\documentstyle[aps,preprint,epsfig]{revtex}

\begin{document}
\draft
 
\title{Heteroskedastic L\'{e}vy flights}
\author{P.\ Santini~\cite{add}}
\address{
Institute of Theoretical Physics, University of Lausanne,
CH--1015 Lausanne, Switzerland}
\maketitle

\begin{abstract}

Truncated L\'{e}vy flights are
random walks in which the arbitrarily large steps
of a L\'{e}vy flight are eliminated.  Since this makes
the variance finite, the central limit theorem applies, and as
time increases the probability distribution of the increments
becomes Gaussian. Here, truncated L\'{e}vy flights with correlated
fluctuations of the variance (heteroskedasticity) are considered.
What makes these processes interesting is the fact that
the crossover to the Gaussian regime may occur
for times considerably larger
than for uncorrelated (or no) variance fluctuations.

These processes may find direct application in the modeling of 
some economic time series.

\end{abstract}

\narrowtext

\section{Introduction}

The ubiquity of the Gaussian probability distribution function (PDF)
in physics and statistics is a
consequence of the central limit theorem (CLT)~\cite{feller},
which states that
the PDF of the sum of $N$ independent, identically distributed (i.i.d.)
stochastic variables whose variance is finite converges to the
Gaussian PDF when $N\rightarrow\infty$.

If the hypothesis of finite variance is relaxed,
a generalized CLT still exists~\cite{levy,kolmo}: the PDF of
the sum belongs to the family of {\it L\'{e}vy} stable distributions,
defined by the characteristic function
(Fourier-transform)
\begin{equation}
L(z)=\exp [{\rm i}mz-\gamma\vert z\vert^{\alpha}(1+{\rm i}\beta
z/\vert z\vert\tan (\alpha\pi /2))]
\end{equation}
with $0< \alpha \le 2$. For $\alpha = 2$ the Gaussian distribution
is recovered, while for $\alpha < 2$ the PDF possesses power-law
tails $L(x) \sim C x^{-(1+\alpha)}$ which make
the variance infinite. In this article, only symmetrically
distributed stochastic variables are considered, for which
$m=\beta =0$.

A {\it L\'{e}vy flight}~\cite{levy,kolmo,report} (LF) is a random walk in which
the step length is chosen from the PDF of Eq.~(1). Since
L\'{e}vy distributions are stable under convolution,
the LF process is self-similar, i.e. the same L\'{e}vy distribution describes
increments over different time scales, provided the increments
are appropriately rescaled.

L\'{e}vy flights appear in various physical
problems~\cite{report,schlesinger}, in particular diffusion, fluid dynamics,
and polymers. However, because of their infinite variance and lack of
a characteristic scale, L\'{e}vy PDFs overestimate the probability of
extreme events when used to model real physical systems, for which
an unavoidable cutoff is always present~\cite{stanleyprl}.
The most direct way to make the variance finite is by means of
{\it truncated L\'{e}vy} (TL) PDFs~\cite{stanleyprl}.
The TL PDF is close to a L\'{e}vy PDF for small argument, but it
contains a sharp~\cite{stanleyprl} or exponential~\cite{koponen}
cutoff in the tails. A {\it truncated L\'{e}vy flight} (TLF) is a random walk in which the step length is chosen from a TL PDF.

The TL PDF belongs
to the basin of attraction of the Gaussian PDF:
for large $N$ the sum of $N$ i.i.d. TL variables is Gaussian-distributed.
For small $N$ the central part of the PDF of the sum has
a L\'{e}vy shape, but the variance and higher moments
are finite. For symmetric TL PDFs the deviation from a Gaussian may be quantified by the value of the
normalized fourth cumulant (kurtosis, $\kappa$)~\cite{feller}.
This is zero for a Gaussian, and positive for a TL PDF.
The crossover to the Gaussian regime
is given by $N \gg N_0 \propto l^\alpha$ with $l$ the cutoff length.
Under this condition, $\kappa$ becomes very small (see Eq.~(20)).

In this article, the TLF stochastic process is generalized
to include a special form of nonlinear dependence of the
increments called {\it heteroskedasticity}.
One can build processes with dependent increments such that
the central part of the PDF of the sum still approximately
behaves as in a L\'{e}vy flight.
What is remarkable and makes these processes
interesting is the fact that the crossover
to the Gaussian PDF may be pushed to values of $N$ which are larger by some
order of magnitudes than if the increments were independent.

\section{L\'{e}vy PDF in economic time series}

Besides being interesting per se, these processes are of direct
of relevance to one of the most noteworthy applications of LF outside
physics, i.e. to the modeling of some financial time series~\cite{nature}.
From a physicist's
perspective, the market is a very good example of a
complex system~\cite{arthurcompl,book1,book2} 
in which the mutual interaction and competition
among a great number of agents (traders
or speculators) with continually adapting strategies, 
together with the influence of unpredictable exogenous factors,
usually produces an intricate out-of-equilibrium
dynamics.

In the classical equilibrium theory of economy
it is assumed that ``equilibrium" values for the prices exist,
satisfying an aggregate,
overall consistency condition (recalling the so called
{\it Nash} equilibria solutions of game theory~\cite{game}).
However, the complex dynamics of market prices does not seem to fit
this classical picture. In particular,
trading volume and price volatility are much higher than expected
from the classical theory~\cite{shiller}. 
From time to time, the market may
display strong movements (crashes or boosts) which cannot always
be understood in terms of rational reactions to incoming new
information~\cite{cutler}. Instead, some of their features
recall the physical concept of self-organized criticality~\cite{sornettesoc}.

At last, 
trading volumes and variances of price increments
change over time~\cite{clark}, and may persist
as low or high for long periods. The existence of such
correlated variance fluctuations
(heteroskedasticity) is difficult to
understand in the framework of classical equilibrium theory,
and few economic models can explain it
(see e.g. Ref.~\onlinecite{haan} where an equilibrium model 
capable to mimic volatility fluctuations of interest rates 
is developed).

A fundamental source of this complex dynamics may 
be found in the inductive, subjective and adaptive
nature of the process leading
agents to formulate the expectations which drive their actions~\cite{arthurcompl,soros,holland,arthur1,morris,muller}, 
an aspect whose fundamental features can be captured by more or
less elaborated game or artificial-life 
models~\cite{arthurcompl,arthur1,blume,arthur2,caldarelli,chopard,lebaronnew,naturelux}. 

Aside the ''microscopic" origin of the complexity of 
financial markets,
another aspect of the problem 
is the phenomenological (''macroscopic") characterization
of this complexity, in particular the study of the properties of
economic time series.
Given the low level of determinism of these
series, the most fruitful description is in terms
of stochastic processes. Martingale processes~\cite{feller}
are the signature of market efficiency. In particular, random walk or
Brownian-motion models have been used for a long time to model the
increments of asset prices~\cite{note1}. This 
might be understood by
the central limit theorem if price changes resulted from the sum of many 
independent random contributions, which would seem a reasonable
assumption.
Indeed, empirical studies of financial time series have revealed
gaussian behavior for long time scales,
typically of the order of several days. However, 
it has been shown~\cite{nature} that
for short time scales 
the central part of the PDF is not Gaussian. It
is well described by a L\'{e}vy distribution, and
therefore suggests an underlying LF rather than random-walk
model.

Non-gaussian scaling
has been found in many economical or financial indeces~\cite{nature,mandelbrot,mantegna91,mantegna98,zayden,cont,bouchaudbook,pictet1}. 
In financial time series, 
scale invariance can be characterized (i.e. the value
of the self-similarity exponent $\alpha$ in Eq.~(1) can be extracted) 
either by comparing the full PDF of price increments over
different time scales, or by studying the time-scale dependence
of some selected properties of the PDF. For example,
the probability of return
\begin{equation}
P_N(x=0) = 
\frac{1}{\pi}\int_{0}^{\infty}(L(z))^{N},
\end{equation}
which depends on the time scale $N$ as $N^{-1/\alpha}$, was used
to extract the value $\alpha \sim 1.40$ in the high-frequency (one-minute)
variations of the Standard {\&} Poor's 500 (S{\&}P500)
index~\cite{nature,mantegna1}.
An alternative quantity which may be used is the so-called Hurst exponent
(see, e.g., Ref.~\onlinecite{bouchaudbook}).

As with other applications of LF,
while L\'{e}vy distributions describe well the central part of the PDF for short times, the power-law tails of these distributions are much fatter
than observed. In particular, the observed variance is finite.
All the previous remarks suggest the TLF as the best candidate to model these series~\cite{bouchaudbook,mantegnaarch}.

Two aspects, however,
remain unexplained : one is the fact that the crossover
to the Gaussian regime occurs at much larger times
than expected from the TLF model (and from any model with
i.i.d. increments as well). For example, the one-minute PDF $P_1$
of the S{\&}P500 index increments has kurtosis
$\kappa_1 = 43$~\cite{nature,mantegnaarch}.
If the PDF at time $N$, $P_N$, were $P_1$
convoluted $N$ times with itself (with kurtosis
$\kappa_N = \kappa_1 /N$), Gaussian behavior would be expected for
$N \gg \kappa_1$ (see Fig.~1).
However, the central part of $P_N$ displays L\'{e}vy behavior
up to at least $N=1000$~\cite{nature}.

The second aspect not accounted for is heteroskedasticity:
even if linear correlations are almost zero,
there are correlations of the squared
increments, i.e. the increments at different times are uncorrelated
yet not independent random variables.
Put another way, one cannot
factorize the joint probability density of the increments at different 
times into the product of reduced densities.

The model proposed in the following can account at the same time for
both these aspects.

\section{Models for heteroskedasticity}
\subsection{Gaussian-type models}

Many efforts have been devoted in the past two decades to
the study of time-varying variance, and various models have been
put forward by econometricians. Roughly, two great classes of models exist :
Auto-Regressive-Conditional-Variance (ARCH) type models
and Stochastic-Variance (SV) type models. 

Models of both classes are usually set up in a Gaussian
framework:
if time is discretized with an elementary time step $\tau$,
the increments at the $k$-th time-step (i.e. at $t=k\tau $) are assumed to be random variables of the form 
\begin{equation}
x_{k}=\mu_{k} + \sqrt{v_k}\epsilon_{k}.
\end{equation}
Here $\mu_{k}$ is the time-varying mean of the stochastic process
($\mu_{k}$ is very small, and can safely be set to zero, for the
short time scales considered here), $v_{k}$ is a random
variable representing the time-varying
variance of the process, and the $\epsilon_{k}$
are independent Gaussian random variables with zero mean and unit
variance, independent of $v_{k}$.

ARCH and SV models differ in the way the $\sigma_{k}$
process is specified : in ARCH-type models~\cite{engle,bollerslev01} 
the variance at time-step $k$ is a deterministic function of the past
squared increments and variances, while in SV-type models~\cite{taylor}
the variance is not completely determined by the past data,
since it contains a random contribution. With suitable
choices of their parameters, both type of models can 
account for the heteroskedasticity and positive kustosis (leptokurtosis) of 
the PDF of financial
series, although usually they fail to capture all aspects of the data~\cite{shephard}.
In particular, it has been shown by numerical simulations in Ref.~\onlinecite{mantegnaarch}
that the simplest ARCH-type models do not yield the L\'{e}vy-type 
scaling of the PDF described above,
since already at short times the value of the scaling exponent is close 
to the gaussian one.

\subsection{General models}

Let us assume
the $x_k$ to be zero-mean
random variables with variance $v_k$, with a symmetric
PDF $P_1^{v_k} (x_k)$
depending on the parameter $v_k$ in an as yet unspecified way.
The parameter $v$ fluctuates in time following a stationary process.
Let
$p_N(v_1,\cdot\cdot ,v_N)$ be the
joint PDF of the
variances at the different times, and let us assume
the joint PDF of increments and variances
\begin{equation}
P_N(x_1,\cdot\cdot ,x_N ; v_1,\cdot\cdot ,v_N) =
p_N(v_1,\cdot\cdot ,v_N)\prod_{i}P_1^{v_i} (x_i),
\end{equation}
i.e. the increments conditional to a certain set
of variances $v_1,\cdot\cdot ,v_N$
are independent variables.
However, $P_N(x_1,\cdot\cdot ,x_N ; v_1,\cdot\cdot ,v_N)$
{\it is not directly observable}.
The object of measurement is the {\it unconditional}
PDF
\begin{equation}
P^{unc}_N(x_1,\cdot\cdot ,x_N) =
\int dv_1\cdot\cdot dv_N P_N(x_1,\cdot\cdot,x_N ; v_1,\cdot\cdot ,v_N),
\end{equation}
which is only factorized if $p_N(v_1,\cdot\cdot,v_N)$ is.

A special case of process with PDF given by Eq.~(4) is
\begin{equation}
x_{k}=\sqrt{v_k} \xi_{k},
\end{equation}
where the $\xi_{k}$
are independent random variables with zero mean and unit
variance, independent of $v_{k}$, and with PDF $P_0 (\xi )$.
In this case,
\begin{equation}
P_1^{v} (x) = \frac{1}{\sqrt{v}} P_0 (\frac{x}{\sqrt{v}}).
\end{equation}
For this special process
the PDF is assumed to
change with time only through a time-varying scale factor.
The process reduces to that of Eq.~(3) if $P_0$ is Gaussian,
however $P_0$ does not necessarily need to be gaussian. For example,
in Ref.~\onlinecite{student} ARCH-type processes with
a $t$-Student $P_0$ are used. 

The characteristic function of $P_1^{v}(x)$ is
\begin{eqnarray}
P_1^{v}(z) &=& \int d x \exp (i z x) P_1^{v} (x) \\
                &\rightarrow& P_0 (\sqrt{v} z).
\end{eqnarray}
Here and in the following $\rightarrow$ indicates the result for the particular
case of a scale-factor-type process, Eq.~(6).
$P_0 (z)$ is the characteristic function of $P_0 (x)$.

$P_1^{v}(z)$ can be used~\cite{feller} to calculate moments and cumulants of $P_1^{v}(x)$
of any order $n$,
\begin{eqnarray}
m_n^v(1) &=& (-{\rm i})^n\frac{{\rm d}^n}{{\rm d}z^n} P_1^{v} (z)\vert_{z=0}
                   \rightarrow v^\frac{n}{2} m_n^0 \\
c_n^v (1) &=& (-{\rm i})^n\frac{{\rm d}^n}{{\rm d}z^n} \log P_1^{v} (z)\vert_{z=0}
                   \rightarrow v^\frac{n}{2} c_n^0,
\end{eqnarray}
$m_n^0$ and $c_n^0$ being moments and cumulants of $P_0$.
The assumed symmetry of $P_1^{v}(x)$ implies that
all its odd-order moments vanish.
The variance $m_2^v (1) = c_2^v (1) = v$.
The normalized cumulants are
\begin{equation}
\eta_n^v (1) = \frac{c_n^v (1)}{v^\frac{n}{2}}
                    \rightarrow c_n^0.
\end{equation}
The kurtosis $\kappa^v (1) = \eta_4^v (1)$ is zero for 
the process of Eq.~(3).

\subsubsection{Unconditional PDF for N=1}.

For the Gaussian-type processes of Eq.~(3), it is easy to extract $p_1(v)$
from a given measured $P_1^{unc}(x)$. Using Eqs.~(4), (5) and (9) its characteristic
function $P_1^{unc}(z)$ is found to be
\begin{equation}
P_1^{unc} (z) = \int {\rm d}v p_1(v) \exp (-\frac{v}{2} z^2).
\end{equation}
Since $p_1(v) = 0$ for $v < 0$ (the variance has to be positive),
Eq.~(13) is a Laplace transform. Setting $z^2/2 = s$,
$P_1^{unc} (\sqrt{2s}) = {\cal L}(p_1(v))$.
Thus, the PDF of the variance
$p_1(v)$ giving an observed $P_1^{unc}(z)$ is
the inverse Laplace transform $p_1(v) = {\cal L}^{-1}(P_1^{unc} (\sqrt{2s}))$.
If $P_1^{unc} (z)$ is a symmetric
L\'{e}vy PDF with index $\alpha$, it follows
that $p_1(v)$ is itself a L\'{e}vy PDF Eq.~(1)
of index $\alpha /2$ and asymmetry $\beta = 1$, $m=0$~\cite{kolmo}.
This latter has an essential singularity at $x=0$.
If $P_1^{unc} (z)$ is a symmetric
TL PDF, the singular behavior close to $x=0$ remains, but the decrease
of $p_1(v)$ for $v\rightarrow\infty$ changes from algebraic to exponential.

For a general process such Laplace-transform relationship is lost. 
By differentiating  $P_1^{unc}(z)$ at $z=0$, moments $m_n^{unc}(1)$ of $P_1^{unc}(x)$
of any order $n$ can be expressed in terms of moments of $P_1^v(x)$ :
\begin{equation}
m_n^{unc}(1) = (-{\rm i})^n\frac{{\rm d}^n}{{\rm d}z^n} P_1^{unc} (z)\vert_{z=0}
             = \langle m_n^{v}(1)\rangle_v,
\end{equation}
where $\langle f(v) \rangle_v = \int {\rm d}v p_1(v) f(v)$ and
$m_n^{v}(1)$ is the $n$-th-order moment of $P_1^v(x)$, Eq.~(10).
The kurtosis of $P_1^{unc}(x)$ is
\begin{equation}
\kappa^{unc}(1) = \frac{m_4^{unc}(1)}{(m_2^{unc}(1))^2} - 3
   \rightarrow \frac{\langle v^2\rangle_v}{\langle v\rangle_v^2}
(\kappa_0 + 3) - 3,
\end{equation}
where $\kappa_0$ is the kurtosis of $P_0$. As is well-known,
a fluctuating variance (for which
$\langle v^2\rangle_v \neq \langle v\rangle_v^2$)
can produce a non-Gaussian PDF
($\kappa^{unc}(1) \neq 0$) even when $P_0$ is Gaussian
($\kappa_0 = 0$).
Thus, a given measured $P_1^{unc} (x)$
may be consistent with many different choices of the couple
($P_1^{v}(x)$ , $p_1(v)$), which explains the existence
of many econometric models. For example, the observed leptokurtic
character of $P_1^{unc} (x)$ may arise either from a leptokurtic
$P_0$ or from fluctuations of the variance or from both effects.

\subsubsection{Unconditional PDF for N $>$ 1}
At time $N>1$
the characteristic function of the unconditional
PDF of the sum, $P_N^{unc} (x)$, $x=\sum_{i=1}^{N} x_i$, is
found from Eq.~(4),
\begin{equation}
P_N^{unc} (z) = \int {\rm d}v_1\cdot\cdot{\rm d}v_N
                 p_N(v_1,\cdot\cdot,v_N)\prod_{i=1}^{N} P_1^{v_i} (z).
\end{equation}
For independent variance fluctuations, $p_N(v_1,\cdot\cdot,v_N) = \prod_{i}p_1(v_i)$,
and $P_N^{unc}(x)$ is simply given by $P_1^{unc}(x)$
convoluted $N$ times with itself. In this case cumulants, including
the variance, scale as $N$. The
kurtosis decreases as $\kappa^{unc}(N) = N^{-1}\kappa^{unc}(1)$.

If fluctuations of the variance are correlated, by differentiating
Eq.~(16)
at $z=0$, moments of $P_N^{unc}(x)$ of any order $n$
may be expressed in terms of
averages of products of
moments of $P_1^v (x)$ taken at different times,
the average being made over
$p_N(v_1,\cdot\cdot ,v_N)$.
The variance scales as $m_2^{unc}(N) = N m_2^{unc}(1)$, just as for uncorrelated (or no)
variance fluctuations. The kurtosis
depends on linear correlations of the variances\cite{eco} :
\begin{eqnarray}
       && \kappa^{unc}(N) =  \kappa^{unc}(1)/N
                       + {\tilde \kappa}^{unc}(N) \\
   && {\tilde \kappa}^{unc}(N) =
                     6(2 + \kappa^{unc}(1))\sum_{d=1}^{N}(1/N-d/N^2)g(d)
                      \nonumber \\
     && g(d) = \frac{\langle v_1 v_{1+d} \rangle_v
                     - \langle v\rangle_v^2}{
                       \langle m_{4}^{v}(1)\rangle_v                                              - \langle v\rangle_v^2}
                  = \frac{\langle\langle x^2_1 x^2_{1+d}\rangle\rangle
                     - \langle\langle x^2\rangle\rangle^2}{
                 \langle\langle x^4\rangle\rangle                                              - \langle\langle x^2\rangle\rangle^2}, \nonumber
\end{eqnarray}
where $\langle..\rangle_v$ is an average over variance fluctuations, and $\langle\langle ..\rangle\rangle$ is an average over fluctuations of
the $x_i$ and the variances.
The normalized two-times autocorrelation
of the squared increments, $g$, 
determines the degree of
persistence of variance fluctuations. 
The simplest
models of the type of Eq.~(3) (such as those considered in
Ref.~\onlinecite{mantegnaarch}) have a positive $g(d) \sim K \exp (-d/d_0)$.
In Ref.~\onlinecite{cont} $g(d)\sim g_0 d^{-0.37}$
is found from the 5-minute increments of the S{\&}P500 index.
In any case, in presence of positive variance correlations, $\kappa^{unc}(N)$ may decrease with $N$ much more slowly
than if $g(d)=0$.
Thus, roughly speaking, the slowing down of the decrease
of $\kappa^{unc}(N)$ pushes
the onset of a Gaussian regime ($\kappa^{unc}(N)\rightarrow 0$)
to much larger values of $N$
than expected from independent (or no)
variance fluctuations. The problem is that for most models,
and in particular those of the type of Eq.~(3),
these heteroskedastic contributions to $P_N^{unc} (x)$
may be inconsistent with L\'{e}vy behavior
close to $x=0$.

To make the variance autocorrelation $g$ explicit in $P_N^{unc} (z)$,
it is convenient to characterize the stochastic process
followed by the variances by its multivariate characteristic function
$p_N(k_1,\cdot\cdot,k_N)$. For independent $v_k$,
$p_N(k_1,\cdot\cdot,k_N) = \prod_{i}p_1(k_i)$, $p_1(k)$ being the Fourier transform of $p_1(v)$. When correlations are present, $p_N(k_1,\cdot\cdot,k_N)$
may be expressed as
\begin{eqnarray}
p_N(k_1,\cdot\cdot ,&k_N&) =
\exp (-A\sum_{i<j}g(i-j) k_ik_j + S) \prod_{i}p_1(k_i) \nonumber \\
&=& \prod_{i}p_1(k_i)(1-A\sum_{i<j}g(i-j)k_ik_j + \cdot\cdot).
\end{eqnarray}
Here $A=\langle m_{4}^{v}(1)\rangle_v- \langle v\rangle_v^2$,
and $S=o((\vert k_1\vert + \cdot \cdot + \vert k_N\vert)^2)$
contains contributions from mixed cumulants of $p_N(v_1,\cdot\cdot,v_N)$ of order
higher than two~\cite{notecum},
e.g. terms of type $k_ik_jk_l$ or $k_ik_jk_lk_s$ with
the time indices $i$, $j$, $l$, $s$ not all equal.
These describe correlations of the variances of higher order than
those described by $g$.
The approximation of putting $S=0$ corresponds to make a Gaussian-like
decoupling of
these high-order correlations into products of linear two-times
($g(d)$) and equal-time correlations (e.g., $\langle v_1^3v_2\rangle
= 3g(1)\langle v^2\rangle + \langle v\rangle \langle v^3\rangle $).

Let us consider, for example, models of the type of Eq.~(3), which
are the most commonly used. For
these models it has been shown (Eq.~(13)) that it is possible to choose $p_1(v)$
such that $P_1^{unc}(x)$ is a TL PDF.
By using Eq.~(18), $g(d)$
may be made explicit in $P_N^{unc} (z)$ , Eq.~(16).
By a few simple
manipulations,
it is possible to sum up all the contributions of $g$ to $P_1^{unc} (z)$
to obtain
\begin{equation}
P_N^{unc} (z)\! =\! P_1^{unc} (z)^N\!\exp (\frac{A z^4}{4}\!\sum_{i<j}g(i-j)
\!+\!{\cal O}(z^6))
\end{equation}
It is peculiar to the processes of Eq.~(3) that
the contributions ${\cal O}(z^6)$ do not depend on $g$, but only
on higher-order correlations ($S$ in Eq.~(18)). More specifically,
mixed cumulants of order $p$ in Eq.~(18) give terms
${\cal O}(z^{2p})$ in the exponent of Eq.~(19).
To produce a L\'{e}vy scaling of $P_N^{unc} (x)$ for small $x$,
$P_N^{unc} (z)$ has to behave as
$[P_1^{unc} (z)]^N\sim\exp (-N\gamma \vert z\vert^\alpha)$ for large $z$. Then,
constraints should be put on variance correlations of any order
to ensure that as $z\rightarrow\infty$ the exponent in Eq.~(19) does not diverge more strongly
than $z^\alpha$. It is also possible that for large $z$
the contributions of variance fluctuations to $P_N^{unc}(z)$ be not
determined by its small $z$ behavior, i.e. by its
moments~\cite{feller}. In turn, this may happen if $p_N(v_1,\cdot\cdot,v_N)$ itself is not uniquely determined by its moments, i.e. $p_N(k_1,\cdot\cdot,k_N)$ has singularities.
Anyway, such non-analytic contributions would not yield a L\'{e}vy scaling
in general.

\section{Truncated L\'{e}vy flights with heteroskedasticity}

It has been shown above that if the elementary PDF $P_1^{v}(x)$ is Gaussian,
L\'{e}vy behavior of $P_N^{unc}(x)$ may
be obtained only
by making {\it ad hoc} assumptions
about $p_1(v)$ and the multivariate structure of variance fluctuations,
which looks somewhat artificial. It is simpler and more intuitive 
to assume, instead, that for some $v$,
$P_1^{v}(x)$ itself is equal or close to a TL PDF. There are many possible
ways to choose how the parameter $v$ should enter.
If a TL PDF with exponential
cutoff at $l\sim \lambda^{-1}$\cite{koponen} is used for $P_1^{v}$,
its variance and kurtosis are, respectively,
\begin{equation}
v = \frac{\gamma \alpha (\alpha -1)\lambda^{\alpha -2}}
            {\vert \cos (\pi\alpha /2)\vert} \hspace{0.2cm} {\rm ,}
\hspace{0.2cm}
\kappa = \frac{(\alpha -3)(\alpha -2)}
            {\lambda^{2}v}.
\end{equation}
If the scaling exponent $\alpha$ is fixed, a time-varying $v$
may occur either through variations of $\gamma$ (these reflect
variations of trading volumes in economic applications), or
variations of the cutoff $\lambda$ or of both.

Although the central part of a superposition of TL PDFs with varying $\gamma$ does not have an exact L\'{e}vy behavior,
this behavior
still exists approximately for many choices of $p_1(v)$,
i.e. one can find a L\'{e}vy PDF with the same $\alpha$ and
an effective $\gamma$ which describes approximately the central part
of the superposition. For example, if $v$ or $\gamma$ have a lognormal PDF
(given that $v$, $\gamma >0$ this is a simple possible candidate PDF), the central part of the superposition is dominated by the $\gamma$ corresponding to the peak
of the lognormal (non-Gaussian region, high kurtosis), while the tail for $\gamma\rightarrow\infty$ (quasi-Gaussian region, small kurtosis)
mostly affects the tails of the superposition. In general, the smaller values
of $\gamma$ determine the small $x$ behavior of $P_1^{unc}(x)$.
On the other hand,
the central part of a superposition of TL PDFs with varying cutoff $\lambda$ keeps an almost exact L\'{e}vy behavior, since
$\lambda$ only affects the tails of the PDF.

In the following we focus on some qualitative features of
TLF with correlated fluctuations of the variance.
We study the simplest conceivable non trivial model,
which nevertheless contains the relevant features
of this type of processes, and has the advantage that an accurate
numerical evaluation of $P_N^{unc}(x)$ can be made without approximations.
The variance is assumed to fluctuate between two
possible values only, $v_a$ and $v_b$, $v_a < v_b$.
Thus, $p_1(v)=p_a\delta (v-v_a)+
(1-p_a)\delta (v-v_b)$. If $p_a \gg 0.5$ one may view  $p_1(v)$
as a stylized lognormal PDF, with $v_a$ representing
the position of the maximum and $v_b$ representing
the tail of the lognormal. For $P_1^{v_a}$ we use a TL PDF
with high kurtosis (strongly non-Gaussian behavior).
$P_1^{v_b}$ should represent a TL PDF with small kurtosis
(quasi-Gaussian behavior). To simplify the model as much as possible,
and minimize the number of parameters,
$P_1^{v_b}$ is assumed
to be Gaussian with variance $v_b$.
Thus, $P_1^{unc}=p_a P_1^{v_a} + (1-p_a) P_1^{v_b}$.

For the stochastic process followed by the variances, the simplest choice
is a Markov chain~\cite{feller}. Thus,
$p_N(v_1,\cdot\cdot,v_N)$ is determined by assigning the probability
$p_a$ at time $N=1$, and
the transition probabilities $p_{aa}$ and $p_{bb}$, i.e.
the probability that if $v=v_a$ (or $v=v_b$) at any time $N$, then $v=v_a$
(or $v=v_b$) at time $N+1$.
If the condition that the Markov chain be stationary
is imposed,
two free parameters, i.e. $p_a$, and $p_{bb}$, are left for the chain.
These are to be added to the 3 parameters ($\alpha$, $\gamma$, $\lambda$ or,
using Eq.~(20), $\alpha$, $v_a$, $\kappa_a$) which fix
$P_1^{v_a}$ and the single parameter ($v_b$) which fixes $P_1^{v_b}$.
Note that for this model $g(d)$, Eq.~(17), decays exponentially
with $d$.

For the purpose of illustration, we try to apply this stylized model
to the S{\&}P500 data of Refs.~\onlinecite{nature,mantegna1}.
Four parameters are fixed by the values
$\alpha=1.4$, $m_2^{unc} (1) \sim 0.016$~\cite{note2},
$\kappa^{unc} (1) = 43$, $P_1^{unc}(x=0) \sim 15$ measured for
the 1-min. increments of the S{\&}P500 index for the period 1984-1989.
A fifth parameter is fixed by the decay rate $d_0$ of $g(d)$.
Although by the Markov chain model it is not possible to directly reproduce the
measured algebraically decaying $g(d)$,
we do not expect this to affect the results
qualitatively as long as the kurtosis is qualitatively well mimicked.
Thus $d_0$ is fixed ($d_0\sim 126$~minutes), roughly, by demanding $\kappa^{unc}(N)$, 
Eq.~(17) (see Fig.~1), to be be equal to unity at the same value 
$N\sim 3700$ which is found from Eq.~(17) if one uses for $g(d)$ the 
same expression $g(d)\sim 0.08 d^{-0.37}$ found in Ref.~\onlinecite{cont}
for the 5-min increments of the S{\&}P500 index future for the period 1991-1995.
Since the time window and the time step of Ref.~\onlinecite{cont}
are not the same of Refs.~\onlinecite{nature,mantegna1}, the value of $d_0$ is
only indicative.

One parameter ($p_a$) remains free, and is fixed at 0.9 (any value
close to one gives similar results).

As it may be seen in the inset of Fig.~1,
the central part of $P_1^{unc} (z)$ is close to a L\'{e}vy PDF.
$P_N^{unc} (z)$ is evaluated up to $N=10000$ by performing $4\cdot 10^7$ simulations of the Markov chain, enough to ensure full convergence of the results. 
These simulations yield a numerical estimation of the multivariate
probability distribution of the variances $p_N(v_1,\cdot\cdot,v_N)$, by which
$P_N^{unc}(z)$ is calculated through Eq.~(16). 
$P_N^{unc}(x)$ is then calculated by a numerical Fourier transform
of $P_N^{unc}(z)$.

The probability of return $P_N^{unc} (x=0)$ is plotted in Fig.~1.
For comparison, $P_N^{unc} (x=0)$ is plotted for
uncorrelated variance fluctuations (in this case $P_N^{unc}(z) =
(p_a P_1^{v_a}(z) + (1-p_a) P_1^{v_b}(z))^N$), and for a TLF with i.i.d.
increments and again $\alpha=1.4$, $m_2 (1) \sim 0.016$,
$\kappa (1) = 43$.
As expected, for the two latter models,
the onset of a Gaussian regime ($P_N^{unc}(x=0) \sim N^{-0.5}$)
occurs as soon as $N \gg \kappa^{unc}(1)$. This would not be in agreement
with observations, since L\'{e}vy scaling, $P_N^{unc}(x=0) \sim N^{-1/\alpha}$, is observed up to at least $N=1000$.
Instead, for the heteroskedastic model the kurtosis decreases much more slowly
(see Fig.~1) and the Gaussian regime
occurs for much larger values of $N$. An
approximate L\'{e}vy scaling persists up to $N \sim 2000$.
In the inset it is shown that such approximate scaling extends to finite
values of $x$ as well,
$P_N^{unc}(x) \sim P_1^{unc}(x N^{-1/\alpha}) N^{-1/\alpha}$~\cite{nature}.
Gaussian scaling is estimated to occur only for $N\agt 30000$.

\section{Conclusions} 

In conclusion, TLF with correlated
fluctuations of the variance (heteroskedasticity) have been considered. 
These processes may be of relevance to the modeling
of some financial time series. 
An explicit numerical calculation has been made by using
for the stochastic process followed by the
variances the simplest conceivable model, i.e. a Markov chain.
Parameters suitable to model the behavior of the
S{\&}P500 stock index have been chosen for illustration.

The central part of the PDF of the increments
during one time step, $P_1^{unc}(x)$,
is close to a L\'{e}vy PDF. What makes these stochastic processes
interesting is the fact that L\'{e}vy scaling of the PDF
may persist for times order of magnitudes larger
than for uncorrelated (or no) variance fluctuations.

It has also been shown that, using the Gaussian-type
models of Eq.~(3), a
L\'{e}vy scaling of the PDF may be obtained only when
quite {\it ad hoc} assumptions
about the multivariate structure of variance fluctuations are made.

\begin{figure}
\epsfig{file=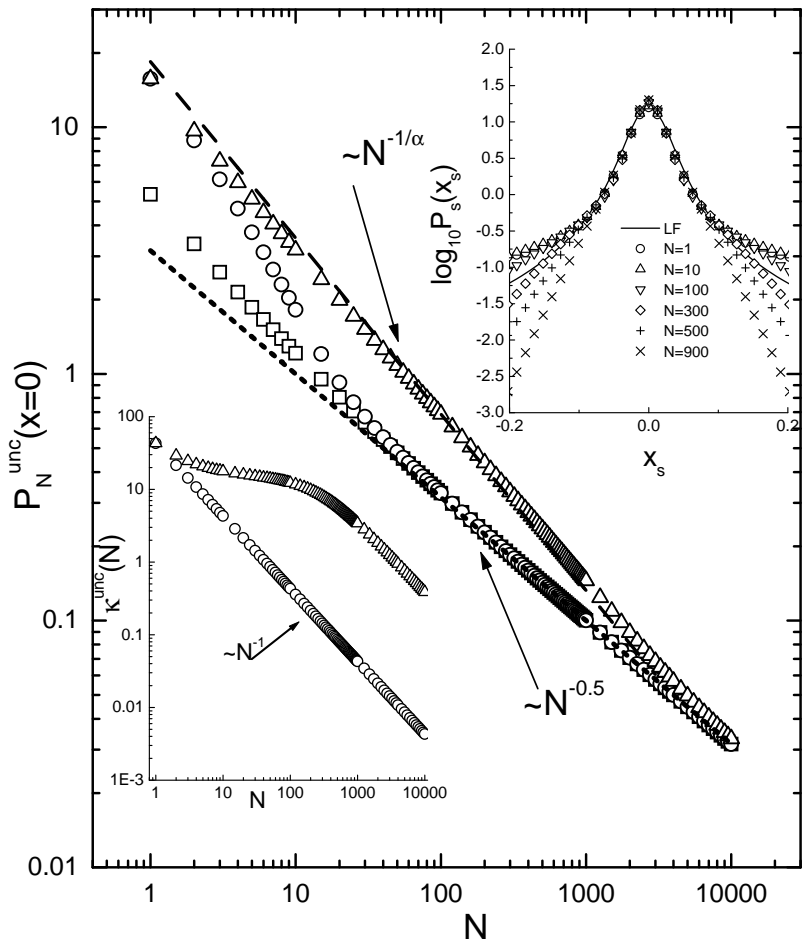,width=.75\textwidth}
\caption{Probability of return {vs} time $N$. Triangles : 
truncated L\'{e}vy flight with Markov-chain heteroskedasticity (HLF)
and parameters $p_a=0.9$, $p_{bb} = 0.9929$, $\alpha = 1.4$,
$v_a = 0.00393$, $\kappa_{a} = 500$, $v_b = 0.1246$, (see text).
$g(N)\sim\exp(-N/126)$. Circles :
i.i.d. variance fluctuations, $g(N>1)=0$. Squares : TLF with
$v=0.016$, $\kappa=43$.
Lower inset : kurtosis {\it vs} time $N$ for the HLF (triangles)
and for the two other non heteroskedastic processes (circles).
Upper inset : scaled PDF,
$P_s(X_s) =  N^{1/\alpha}P_N^{unc}(X_s)$, $X_s = x N^{1/\alpha}$
for the HLF with various values of $N$ compared with a L\'{e}vy PDF
with $\alpha = 1.4$ and $\gamma = 0.0037$.}

\end{figure}

\begin{references}
\bibitem{notecum} If all the moments of $p_1(v)$ are finite (this is true
for the processes of Eq.~(3) because $p_1(v)$ decays exponentially at large $v$,
see discussion after Eq.~(13)),
because of H\"{o}lder's inequality~\cite{feller}, all mixed moments
and cumulants of $p_N(v_1,\cdot\cdot,v_N)$ are finite, too.
\bibitem{note2} Below 20 minutes weak linear
of the increments yield a superdiffusive behavior (i.e. $m_2^{unc} (N)
\propto N^{s}$, $s > 1$), which cannot
be included in the present model.
The value $m_2^{unc} (1) = 0.016$ is of the order of
the average $m_2^{unc} (N)/N$ measured for $N < 1000$~\cite{mantegna1}.
\end{references}
\end{document}